\documentclass[aip, jcp, amsmath, amsthm, amssymb, citeautoscript, reprint]{revtex4-2}
\usepackage{amsmath, amsthm, amssymb}
\usepackage{graphicx, float, xcolor, pgf}
\usepackage{physics, siunitx}
\usepackage[caption=false, justification=centerlast]{subfig}
\usepackage{soul}
\usepackage{gensymb}
\usepackage{hyperref}
\hypersetup{
    pdfauthor={Devansh Sharma and Amartya Bose},
    pdftitle={A Non-Hermitian State-to-State Analysis of Transport in Aggregates with Multiple Endpoints},
    colorlinks=true,
}

\allowdisplaybreaks
\begin{document}
\title{A Non-Hermitian State-to-State Analysis of Transport in Aggregates with Multiple Endpoints}
\author{Devansh Sharma}
\affiliation{Department of Chemical Sciences, Tata Institute of Fundamental Research, Mumbai 400005, India}
\author{Amartya Bose}
\email{amartya.bose@tifr.res.in}
\affiliation{Department of Chemical Sciences, Tata Institute of Fundamental Research, Mumbai 400005, India}
\begin{abstract}
Efficiency of quantum transport through aggregates with multiple end-points or traps proves to be an emergent and a highly non-equilibrium phenomenon. We present a numerically exact approach for computing the emergent time scale and amount of extraction specific to particular traps leveraging a non-Hermitian generalization of the recently introduced state-to-state transport analysis~[Bose and Walters, J. Chem. Theory Comput. 2023, \textbf{19}, 15, 4828–4836]. This method is able to simultaneously account for the coupling between  various sites, the many-body effects brought in by the vibrations and environment held at a non-zero temperature, and the local extraction processes described by non-Hermitian terms in the Hamiltonian. In fact, our non-Hermitian state-to-state analysis goes beyond merely providing an emergent loss time-scale. It can parse the entire dynamics into the constituent internal transport pathways and loss to environment. We demonstrate this method using examples of an exciton transport in a lossy polaritonic cavity. The loss at the cavity and the extraction of the exciton from a terminal molecule provide competing mechanisms that our method helps to unravel, revealing extremely interesting non-intuitive physics. This non-Hermitian state-to-state analysis technique contributes an important link in understanding and elucidating the routes of transport in open quantum systems.
\end{abstract}
\maketitle

Various molecular aggregates function as wires, transferring charges and
excitation from one end to the other. A prime example of such transport
happening in nature are the so-called light-harvesting antenna complexes that
absorb solar photons converting them into excitons, which are then shuttled to the
reaction center where further reactions take place. These natural systems often
have unprecedented efficiencies, which have been a subject of extensive
study~\cite{engelEvidenceWavelikeEnergy2007,
ishizakiTheoreticalExaminationQuantum2009, duanNatureDoesNot2017,
duanQuantumCoherentEnergy2022}. It is important to be able to quantify and
simulate the efficiency of transport in these complex aggregates. The thermal
environment modulates these transfers in a non-perturbative manner.
Consequently, advanced wave function-based techniques like the density matrix
renormalization group (DMRG)~\cite{whiteDensityMatrixFormulation1992,
schollwockDensitymatrixRenormalizationGroup2005,
schollwockDensitymatrixRenormalizationGroup2011} or the multi-configuration
time-dependent Hartree
(MCTDH)~\cite{meyerMulticonfigurationalTimedependentHartree1990,
beckMulticonfigurationTimedependentHartree2000} cannot be used effectively since
they are incapable of capturing the continuum manifold of environmental states
that are thermally accessible.

Simulations involving reduced density matrix provide a lucrative approach to
understanding such non-equilibrium transport. Heijs \textit{et
al.}~\cite{heijsTrappingTimeStatistics2004} and Cao \&
Silbey~\cite{caoOptimizationExcitonTrapping2009} have explored the relation
between trapping time and efficiency, providing a classical kinetic picture.
Efficiency of quantum transport has been studied using
approximate~\cite{uchiyamaEnvironmentalEngineeringQuantum2018,
dijkstraEfficientLongdistanceEnergy2019} and numerically exact
methods~\cite{kreisbeckHighPerformanceSolutionHierarchical2011}. Sener
\textit{et al.}~\cite{senerRobustnessOptimalityLight2002} and others have
explored the robustness of photosynthetic transport. Approximate methods are
often plagued by \textit{ad hoc} assumptions that may fail for a particular
system. In this context, numerically exact simulations using methods like
hierarchical equations of motion (HEOM)~\cite{tanimuraTimeEvolutionQuantum1989,
tanimuraReducedHierarchicalEquations2014, tanimuraNumericallyExactApproach2020,
xuTamingQuantumNoise2022, ikedaGeneralizationHierarchicalEquations2020} or the
quasi-adiabatic propagator path integral
(QuAPI)~\cite{makriTensorPropagatorIterativeI1995,
makriTensorPropagatorIterativeII1995, makriLongtimeQuantumSimulation1996,
strathearnEfficientNonMarkovianQuantum2018, bosePairwiseConnectedTensor2022,
boseMultisiteDecompositionTensor2022, boseQuantumCorrelationFunctions2023} are
extremely useful. They however require a full description of the end-point from
which the extraction happens, the sites to which the extraction happens, and
their thermal environments to be able to predict the efficiency of the transport.
Many transport aggregates may even have more than one end-point or trap site. In
such cases, this already challenging parameterization requirement is followed by
an exponential growth of complexity due to a growth of the system Hilbert space
with every extra trap site. Additionally, many processes like spontaneous
emission from an excited state, or loss from a leaky cavity in case of a
polaritonic system cannot be simply expressed as a well-characterized harmonic
bath. These processes are naturally defined in terms of empirical time-scales.
The grand challenge, therefore, is simulating the dynamics under empirical loss
processes using numerically exact methods and then calculating the individual
efficiencies of the different traps in a multi-trap transport process.

The recently developed path integral Lindblad dynamics framework
(PILD)~\cite{boseIncorporationEmpiricalGain2024} allows for a combination of
Lindblad master equation to incorporate the loss or gain processes and
numerically rigorous path integrals to account for the effects of the thermal
environment. PILD has been used to study the effect of loss processes on exciton
transport dynamics in Fenna-Matthews-Olson
complex~\cite{boseIncorporationEmpiricalGain2024} as well as on linear spectra
of chiral aggregates~\cite{sharmaImpactLossMechanisms2024}. In many cases, the
use of non-Hermitian descriptions of the system coupled with path integrals to
incorporate the dissipative environment provides an alternative route to study
the dynamics~\cite{palmNonperturbativeEnvironmentalInfluence2017}. While the
non-unitarity of the propagators make it inherently unsuited for spectra
represented by correlation functions that couple the ground and the excited
state manifolds, it can be enough for the study of dynamics.

Understanding the time-scale of transport in aggregates with multiple monomers,
while important, turns out to be quite challenging. Imagine an aggregate where,
in the simplest case, the exciton or other quantum particle is extracted from
one of the molecular sites, called the ``trap'' site, with a ``local''
time-scale of $T$ time 
units. This $T$ units, which is the time-scale of extraction
from an isolated site, is not the time-scale which one would actually observe in
the aggregate. From the site of exciton injection, it would have traveled to all
accessible sites and the fraction that is on the trap site would get extracted.
This combination of non-equilibrium processes gives rise to an emergent
time-scale $\tau$ for the extraction as observed, when a particular site is
initially excited. It is this emergent time-scale, $\tau$, that is of interest
to us. The problem becomes even more challenging when there are multiple trap
sites with different local extraction time-scales, $T_j$, splitting the single
exciton in different proportions. A relevant example of this multi-trap
transport turns out to be a single channel transport aggregate coupled to a
lossy Fabry-P\'erot cavity --- the loss at the cavity providing a different
trapping channel which, while useless to the transport, needs to be
incorporated. The question then becomes one of assigning a site-specific
emergent time-scale $\tau_j$ and also understanding how much of the exciton is
extracted ($L_j^\infty$) from each such site.

In this work, we address this question of calculating the trap-specific
transport efficiency corresponding to a multi-trap transport aggregate in terms
of the emergent site-specific timescale, $\tau_j$, and exciton extraction,
$L_j^\infty$, by generalizing the recently derived state-to-state
analysis~\cite{boseImpactSolventStatetoState2023} for Hamiltonians with on-site
non-Hermiticities, describing the rates of leakage from the different traps.
This allows us to partition the total leakage of the system into the different
sites enabling us to build an intuitive picture for the efficiency of transport
for a particular non-equilibrium initial condition without any \textit{ad hoc}
approximations.

Consider the following Hamiltonian that describes an open quantum system:
\begin{align}
    \hat{H} &= \hat{H}_0 + \hat{H}_\text{env}
\end{align}
where $\hat{H}_0$ describes the system and $\hat{H}_\text{env}$ describes the
environment and its interaction with the system. For concreteness, let us assume
that the system is described by a Frenkel Hamiltonian where each of the basis
vectors $\ket{j}$ represents the quantum particle (charge or excitation) is on
the $j$th site or molecule and every other site being empty. The system, then,
is generically described by the non-Hermitian operator,
\begin{align}
    \hat{H}_0 &= \sum_j\epsilon_j\dyad{j} + \sum_{j<k} h_{jk}\left(\dyad{j}{k} + \dyad{k}{j}\right),
\end{align}
where $h_{jk}\in\mathbb{R}$ is the coupling or hopping parameter between the
$j$th and the $k$th sites, and $\epsilon_j\in\mathbb{C}$ is the energy of the
system when the particle is on the $j$th site with the corresponding lifetime.
The real part of $\epsilon_j$ represents the site energy while the imaginary part of
$\epsilon_j$, wherever non-zero, represents the rate of loss from the $j$th site
corresponding to the ``local" decay time $T_j$ with
$\Im\left(\epsilon_j\right)=-\pi\hbar\,/\,T_j$.

Some or all of the sites are individually coupled to thermal environments
\begin{align}
    \hat{H}_\text{env} &= \sum_j \hat{H}_\text{env}^{(j)}\\
    \hat{H}_\text{env}^{(j)} &= \sum_b \frac{p_{jb}^2}{2} + \frac{1}{2}\omega_{jb}^2 x_{jb}^2 - c_{jb}x_{jb}\hat{s}_j
\end{align}
where the bath on the $j$th site is coupled to the system through
the system operator $\hat{s}_j$. Each bath of harmonic oscillators is
characterized by the oscillators' frequencies $\omega_{jb}$ and their
corresponding couplings $c_{jb}$. These are related to the spectral density,
\begin{align}
    J_j(\omega) &= \frac{\pi}{2}\sum_b \frac{c^2_{jb}}{\omega_{jb}}\delta(\omega-\omega_{jb}),
\end{align}
which can be estimated using molecular dynamics
simulations~\cite{makriLinearResponseApproximation1999,
boseZerocostCorrectionsInfluence2022} or directly from experiments.

After an initial Frank-Condon excitation of a molecule in the aggregate, the
state of the system can be written in a separable form $\rho(0) =
\tilde\rho(0)\otimes e^{-\beta H_\text{env}}/{Z}$. The time-evolved reduced
density matrix can be written in terms of path
integrals~\cite{makriTensorPropagatorIterativeI1995,
makriTensorPropagatorIterativeII1995} (augmented for non-Hermitian
systems~\cite{palmNonperturbativeEnvironmentalInfluence2017}) as,
\begin{align}
    \mel{s_N^+}{\tilde\rho(N\Delta t)}{s_N^-} &= \sum_{s_j^\pm}\mel{s_N^+}{U}{s_{N-1}^+}\mel{s_{N-1}^+}{U}{s_{N-2}^+}\cdots\nonumber\\
    &\times\mel{s_1^+}{U}{s_0^+}\mel{s_0^+}{\rho(0)}{s_0^-}\mel{s_0^-}{\bar{U}}{s_1^-}\nonumber\\
    &\times\cdots\mel{s_{N-1}^-}{\bar{U}}{s_N^-}\times F\left[\left\{s_j^\pm\right\}\right]
\end{align}
where $U = \exp(-i \hat{H}_0 \Delta t/\hbar)$ is the forward propagator, and
$\bar{U} = \exp(i \hat{H}_0^\dag \Delta t / \hbar)$ is the backward propagator.
Notice that the non-Hermiticity of the system ($\hat{H}_0\ne\hat{H}_0^\dag$) is
taken into account in the definition of the backward propagator. In the path
integral, the state of the system at the $j$th time point is $s_j^\pm$. The
Feynman-Vernon influence functional~\cite{feynmanTheoryGeneralQuantum1963},
$F\left[\left\{s_j^\pm\right\}\right]$, captures the system-environment
interaction and makes the dynamics non-Markovian. It is dependent upon the
spectral density~\cite{boseZerocostCorrectionsInfluence2022}. In a condensed
phase medium, the decay of the memory with time allows for a truncation of the
memory and iteration beyond this memory length. However, the cost of simulation
still increases exponentially within the memory length. Here, we use the
time-evolving matrix product operator (TEMPO)
algorithm~\cite{strathearnEfficientNonMarkovianQuantum2018} adapted for
non-Hermitian systems to do the simulations efficiently. This is implemented in
the recently released \texttt{QuantumDynamics.jl}
package~\cite{boseQuantumDynamicsJlModular2023}.

\begin{figure}
    \centering
    \includegraphics{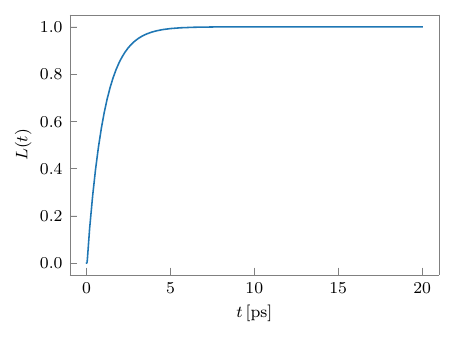}
    \caption{Excitation loss $L(t)$ in the excitonic trimer with a single trap starting with an initial excitation $\tilde{\rho}(0)=\dyad{1}$ and $T_3=\SI{0.3}{\ps}$.}
    \label{fig:loss_trimer}
\end{figure}

Now, because of the non-Hermiticity of $\hat{H}_0$ and the consequent
non-unitarity of the time-evolution, $\Tr\left[\tilde\rho(t)\right]\le 1$. In
fact, the trace is a monotonically decreasing quantity and for the single
excitation subspace, the quantity $L(t) = 1 - \Tr\left[\tilde\rho(t)\right]$ is
the amount of excitation that has leaked out of the system. As a simple example,
consider excitation transport in an excitonic trimer of identical monomers, with
a constant nearest-neighbor electronic coupling $h_{jk} = h =
\SI{-181.5}{\per\cm}\,\delta_{k, j+1}$ and an exciton drain on the third monomer
with a decay time of $T_3=\SI{0.3}{ps}$. This coupling is representative of
typical electronic couplings in bacteriochlorophyll
chains~\cite{tretiakBacteriochlorophyllCarotenoidExcitonic2000,
freibergExcitonicPolaronsQuasionedimensional2009,
boseAllModeQuantumClassical2020}. The vibronic couplings and the solvent
interactions associated with each monomer is modeled as
\begin{align}
   J(\omega) = \frac{\pi}{2}\,\hbar\,\xi\omega\exp\left(-\omega/\omega_c\right),
\end{align}
where $\xi=0.121$ and $\omega_c=\SI{900}{\per\cm}$ corresponding to a
reorganization energy $\lambda_0=\SI{218}{\per\cm}$. The temperature is set to
$\SI{300}{\kelvin}$. (This particular system will be used in different contexts
throughout the paper. For simplicity, we will refer to it as the ``excitonic
trimer.'') The transport starts with excitation on the first monomer,
$\tilde\rho(0)=\dyad{1}$. The loss $L(t)$ for the excitonic trimer is shown in
Figure~\ref{fig:loss_trimer}. The emergent time-scale of loss, obtained with the
model fit $L(t)=1 - \exp(-t/\tau)$, is $\tau = \SI{1.04}{\ps}$, which is
significantly longer than the local loss time-scale of $T_3=\SI{0.3}{\ps}$.

If we were interested in the transport through a system with only one trap site,
then considering the dynamics of $L(t)$ as demonstrated would be adequate.
However, we want to generalize this to molecular aggregates with multiple traps,
where we would like to extract the trap-specific efficiency. The primary
complication that arises with the previous argument of the loss being the change
in the trace of the density matrix applied to this case is that the loss can now
happen through more than one site. How do we figure out the partitioning of this
total loss into the constituent single site losses? To achieve this, we
generalize the concept of the state-to-state
transfer~\cite{boseImpactSolventStatetoState2023} to account for non-Hermitian
systems.

Consider the rate of change of population of the $j$th site,
\begin{align}
    \pdv{P_j}{t} &= \pdv{t}\Tr_\text{sys-env}\big[\,\rho(t) \dyad{j}\big],\label{eq:deriv_Pj_1}
\end{align}
where $\rho(t)$ is the time-evolved density matrix in the full
system-environment Hilbert space. Because the Hamiltonian is non-Hermitian, one
can write the quantum Liouville equation as
\begin{align}
    \pdv{\rho}{t} &= -\frac{i}{\hbar}\left(\hat{H}\rho - \rho\hat{H}^\dag\right).\label{eq:qLE}
\end{align}
Consequently, Eq.~\ref{eq:deriv_Pj_1} can be written as
\begin{align}
    \pdv{P_j}{t} &= -\frac{i}{\hbar}\Tr_\text{sys-env}\left[\left(\hat{H}\rho - \rho\hat{H}^\dag\right)\dyad{j}\right]\\
    &= \frac{i}{\hbar}\Tr_\text{sys}\left[\tilde\rho(t)\left(\hat{H}_0^\dag\dyad{j} - \dyad{j}\hat{H}_0\right)\right],
\end{align}
using the fact that the projector, $\dyad{j}$, commutes with
$\hat{H}_\text{env}$. Expanding the trace, we can now partition this flux into
the source sites:
\begin{align}
    \pdv{P_j}{t} &= \frac{i}{\hbar}\sum_k\left(\mel{j}{\tilde\rho(t)}{k}\mel{k}{\hat{H}_0^\dag}{j}\right.\nonumber\\
    &- \left. \mel{j}{\hat{H}_0}{k}\mel{k}{\tilde\rho(t)}{j}\right)\label{eq:rate_partition}
\end{align}
For every $k$, the summand expresses the instantaneous rate of transfer from $k$
to $j$ and the total rate of change of the population of the $j$th site is given
as a sum over all $k$ as expressed in Eq.~\ref{eq:rate_partition}. Integrating
over time, one gets the total partitioned transfer between the $k$th and the $j$th
sites till time $t$:
\begin{align}
    P_{j\leftarrow k}(t) &= \frac{i}{\hbar}\int_0^t\dd{t'} \left(\mel{k}{\hat{H}_0^\dag}{j} - \mel{j}{\hat{H}_0}{k}\right)\Re\mel{j}{\tilde\rho(t')}{k}\nonumber\\
    &- \frac{1}{\hbar}\int_0^t\dd{t'} \left(\mel{k}{\hat{H}_0^\dag}{j} + \mel{j}{\hat{H}_0}{k}\right)\Im\mel{j}{\tilde\rho(t')}{k}
\end{align}
This equation is completely general. For Hermitian systems, only the second term
survives, and the expression reduces to the previously derived expression for
state-to-state transport~\cite{boseImpactSolventStatetoState2023,
boseImpactSpatialInhomogeneity2023}. The extra physics corresponding to the
losses are all incorporated in the first term. Because non-Hermiticity is
limited to the diagonal on-site terms, note that the site-to-site transport
($j\ne k$) is still governed only by the second term, which becomes
$P_{j\leftarrow k}(t) = -\frac{2}{\hbar}\int_0^t \dd{t'}\mel{j}{\hat{H}_0}{k}
\Im\mel{j}{\tilde\rho(t')}{k}$. For the self transfer terms ($j=k$), the second
term becomes zero because the diagonal elements of the density matrix are real.
However the first term is non-zero with $P_{j\leftarrow j}(t) =
\frac{2}{\hbar}\int_0^t \dd{t'}
\Im\mel{j}{\hat{H}_0}{j}\Re\mel{j}{\tilde\rho(t)}{j}$. Notice that if a site has
a loss term, $\Im\mel{j}{\hat{H}_0}{j}<0$ and $P_{j\leftarrow j}(t)<0$
symbolizing a loss from the site into the environment. Thus, the loss through
the $j$th site, $L_j(t)$, is identified with $P_{j\leftarrow j}(t)$, and the total
loss can be expressed as a sum over all the site-based losses. To obtain the
emergent time-scale, this can now be fit to the following model (assuming a
single exponential decay),
\begin{align}
    L_j(t) &= L_j^\infty\left(1-\exp\left(-t/\tau_j\right)\right)\label{eq:fit}
\end{align}
where $\tau_j$ is the emergent decay time for site $j$ and $L_j^\infty$ is the
net loss through that site. These quantities will now be used to characterize
efficiency of transport with a single time-scale. Notice that $L_j(0) = 0$ and
$\lim_{t\to\infty}L_j(t)=L_j^\infty$.

\begin{figure}
    \centering
    \includegraphics{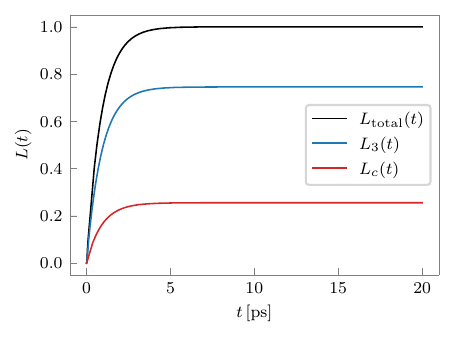}
    \caption{Excitation loss $L(t)$ in the polaritonic trimer starting with an initial excitation $\tilde\rho(0)=\dyad{1}$ through an exciton drain of decay time $T_3=\SI{0.3}{ps}$ coupled to a leaky optical cavity of lifetime $T_c=\SI{0.6}{ps}$.}
    \label{fig:loss_trimer_cav}
\end{figure}

In order to demonstrate this method, we consider a polaritonic system --- the
same excitonic trimer of identical monomers as discussed earlier now coupled to a leaky Fabry-P\'erot cavity of
energy $\hbar\omega_c$ with a coupling strength of $\Omega=\SI{181.5}{\per\cm}$.
The cavity mode energy is set in resonance with the monomer's vertical
(Frank-Condon) excitation energy. The new system Hamiltonian $\hat{H}_{0}^{'}$
is given by
\begin{align}
    \hat{H}_{0}^{'} &= \hat{H}_{0} + \hbar\left(\omega_c-i\pi/T_c\right)\dyad{c} \nonumber \\ &\hspace{1em}+ \sum_j\Omega\left(\dyad{j}{c}+\dyad{c}{j}\right),
\end{align}
where $\ket{c}$ is the cavity mode. The cavity mode has a lifetime $T_c$ of
$\SI{0.6}{\ps}$, which is slightly on the longer
side~\cite{delpinoTensorNetworkSimulation2018}. Note that the cavity is not
associated with any bath. (This system will be called the ``polaritonic
trimer.'') The initial excitation is again set to be on the first monomer,
$\tilde\rho(0)=\dyad{1}$. The loss from the third monomer (exciton drain) and the
cavity are shown in Figure~\ref{fig:loss_trimer_cav}. As mentioned, the total
loss, $L(t)$, is the sum of losses from the exciton drain, $L_3(t)$, and the cavity,
$L_c(t)$. The emergent loss time-scales for the exciton drain, $\tau_3$, and
cavity, $\tau_c$, are $\SI{0.89}{\ps}$ and $\SI{0.90}{\ps}$, respectively. In this
case, a large majority (around $74\%$) of the extraction happens through the
molecular drain site. 

\begin{figure}
    \centering
    \hspace*{-0.5cm}
    \subfloat[Flow from non-leaky sites into exciton drain]{\includegraphics{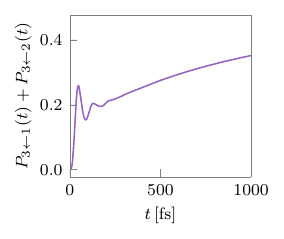}}
    \subfloat[Flow from non-leaky sites into cavity]{\includegraphics{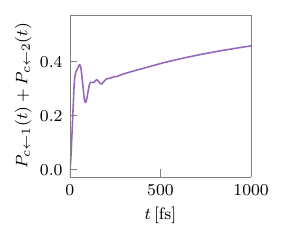}}

    \hspace*{-0.5cm}
    \subfloat[Flow from cavity to exciton drain]{\includegraphics{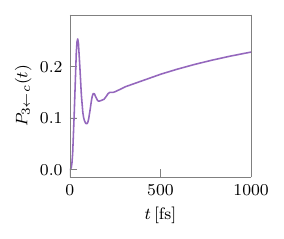}}
    \subfloat[Population of leaky sites]{\includegraphics{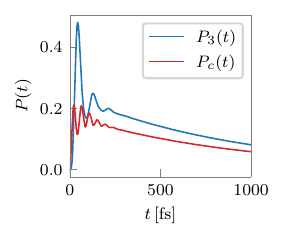}}
    
    \caption{A state-to-state analysis of excitation flows into the trap sites and their population dynamics for the polaritonic trimer.}
    \label{fig:state}
\end{figure}

It is pertinent at this point to ask how exactly is the excitation flowing
through the system to reach these leaky sites. A state-to-state analysis is
presented in Figure~\ref{fig:state} to show the excitation flows to the leaky
sites along with their populations. Notice that at very short times, the flow
from the non-leaky sites, $\ket{1}$ and $\ket{2}$, into the cavity $\ket{c}$
(Figure~\ref{fig:state}~(b)) is more than that into the exciton drain site
$\ket{3}$ (Figure~\ref{fig:state}~(a)). This is because both $\ket{1}$ and
$\ket{2}$ transfer population to the cavity site, whereas only $\ket{2}$
transfers to $\ket{3}$ owing to the nearest-neighbor nature of the excitonic
Hamiltonian. However, very soon the transfer from $\ket{2}$ to $\ket{3}$ catches
up. What is very interesting is that there is a non-insignificant flow from the
cavity site $\ket{c}$ to the exciton drain site $\ket{3}$ from very early
times (Figure~\ref{fig:state}~(c)). All of this along with the leakages from
$\ket{c}$ and $\ket{3}$ come together to give the total population dynamics that
we can see in Figure~\ref{fig:state}~(d). Notice that the initial build up of
population is higher in $\ket{3}$ than in the cavity even though the leakage on
the cavity in this case is slower than that of the monomer.

\begin{figure}
    \centering
    \includegraphics{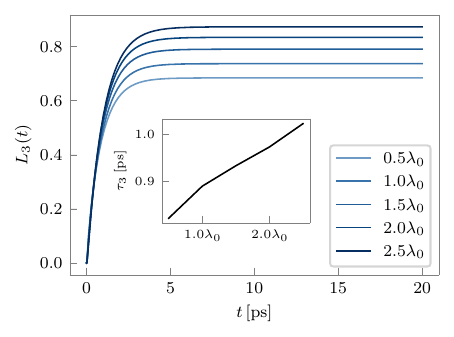}
    \caption{Loss through exciton drain $L_3(t)$ with increasing bath reorganization energies, $\lambda$, and the emergent time-scales $\tau_3$ (inset) for the polaritonic trimer.}
    \label{fig:loss_lambda}
\end{figure}

\begin{figure}
    \centering
    \includegraphics[width=2.25in]{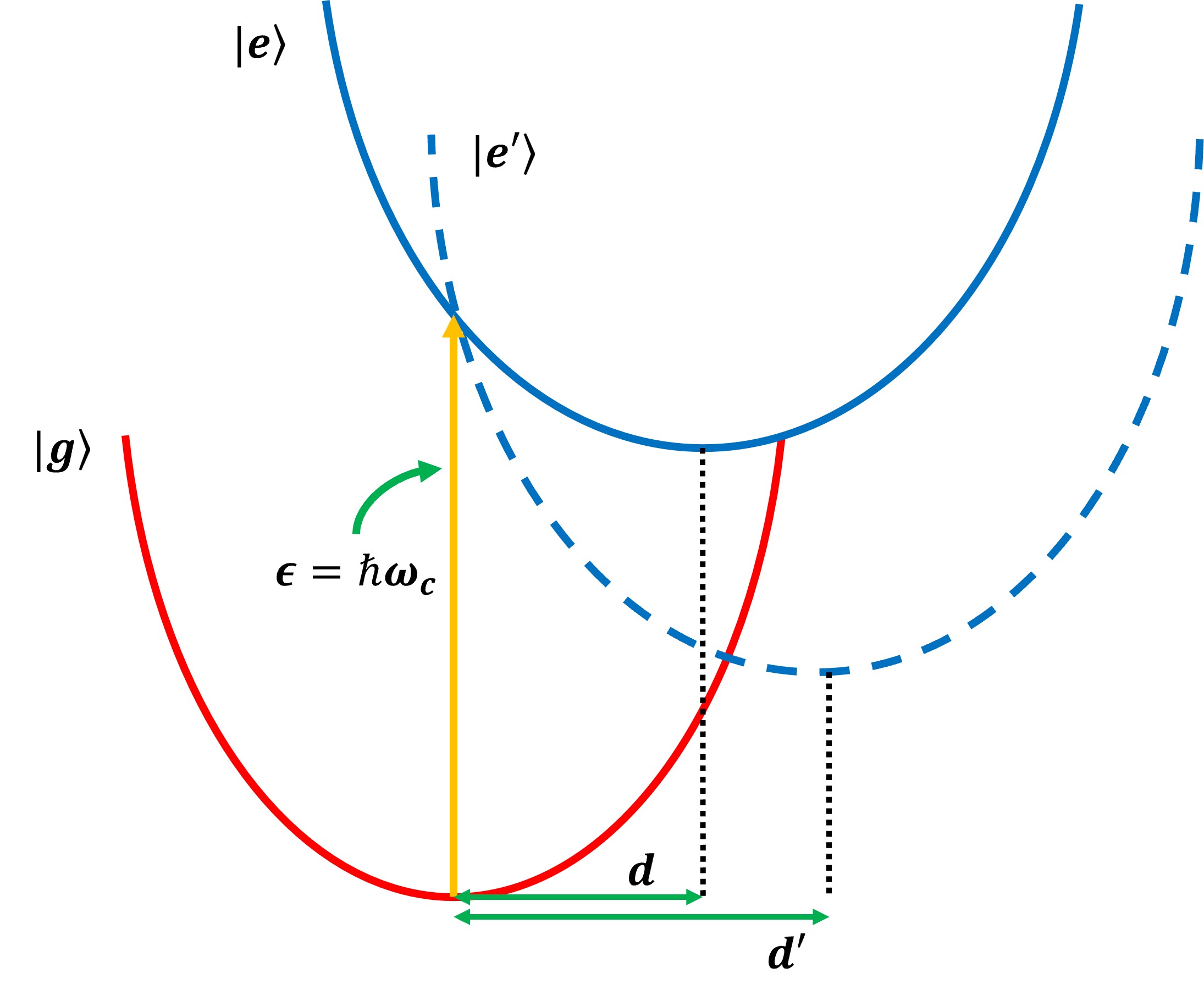}
    \caption{Schematic depicting shift in excited energy surfaces for two baths differing in their reorganization energies. The $\ket{e'}$ surface corresponds to the higher reorganization energy bath than the $\ket{e}$ surface leading to larger displacement $d'>d$.}
    \label{fig:pes}
\end{figure}

\begin{figure}
    \centering
    \includegraphics{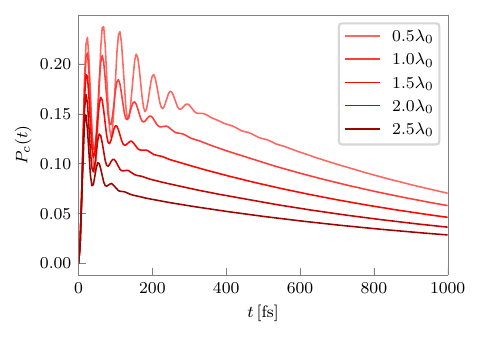}
    \caption{Population of the cavity site, $P_c(t)$, for different reorganization energies, $\lambda$, in the polaritonic trimer.}
    \label{fig:pop_cavity}
\end{figure}

Now let us see what happens when the reorganization energy of the monomeric
environment increases. The losses through the exciton drain $L_3(t)$ for the
same system Hamiltonian parameters but with different bath reorganization
energies are presented in Figure~\ref{fig:loss_lambda}. It is clear that the
amount of extraction through the third monomer ($L_3^\infty$) increases with
the reorganization energy. As a corollary, this would imply that the amount of
leakage through the cavity is decreasing. To understand this better, recall that
the cavity energy is set to be resonant with the Frank-Condon (vertical)
excitation energy of the monomer. The higher the reorganization energy, the
greater is the shift between the ground and excited states of the monomer
because the reorganization energy per ``mode'' is proportional to the Huang-Rhys
factor and consequently the relative displacement of the surfaces. In
Figure~\ref{fig:pes}, a schematic is shown for the current situation. Suppose
that $\ket{e}$ is the excited state surface for the lower reorganization energy
environment and $\ket{e'}$ is the one for the higher reorganization energy
environment. Notice that $d'>d$. Now, to ensure that the vertical excitation
energy remains constant, the surface of $\ket{e'}$ has to be stabilized more.
This means that the minimum of the excited state potential energy surface gets
increasingly detuned from the cavity energy, hampering the transport into the
cavity mode from the monomers. In Figure~\ref{fig:pop_cavity}, we show the
population of the cavity site. Notice that the accumulation of excitation in the
cavity mode decreases with increasing reorganization energy on the monomers.
This verifies the argument presented.

Having discussed the increase in $L_3^\infty$ as a function of $\lambda$, we
turn our attention to the emergent time-scales. From the inset plot in
Figure~\ref{fig:loss_lambda}, we see that $\tau_3$ increases monotonically with
$\lambda$. The initial slopes of the loss curves in Figure~\ref{fig:loss_lambda}
are identical. This coupled with the fact that the $L_3^\infty$ values are
increasing with the reorganization energy, means that the time-scale of exciton
leakage through the third monomer, $\tau_3$, also increases. This becomes
obvious if we remember that according to Equation~\ref{eq:fit}, $\tau_j$ is
effectively related to the half-life of the particle leaking out of the $j$th
monomer.

It should be noted here that it is not necessary that the loss dynamics would
have a single time-scale. In fact, the exponential fit to the loss function,
Equation~\ref{eq:fit}, begins to fail when the reorganization energy of the
environment decreases beyond a certain limit due to a preponderance of
transients in the dynamics. This parallels failure of rate theory to predict the
actual dynamics when transients are
important~\cite{boseNonequilibriumReactiveFlux2017}. The present non-Hermitian
state-to-state analysis method can be used even when this happens. It is just
that we would not be able to talk in terms of the emergent time-scales and would
have to directly study the loss curves similar to the ones in
Figure~\ref{fig:loss_lambda}.

 \begin{figure}
     \centering
     \hspace*{-0.5cm}
     \subfloat[Net loss through exciton drain]{\includegraphics{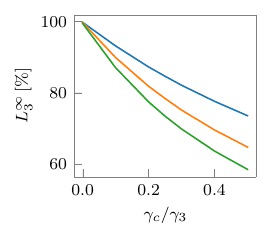}}
     \subfloat[Emergent time-scale]{\includegraphics{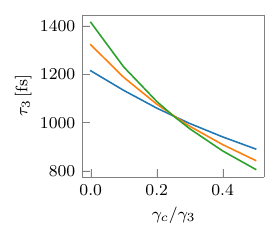}}

     \hspace*{-0.5cm}
     \subfloat[Population of trap site, $P_3(t)$, for $\gamma_c/\gamma_3=0.25$]{\includegraphics{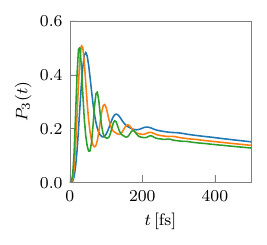}}
     \subfloat[Population of cavity mode, $P_c(t)$, for $\gamma_c/\gamma_3=0.25$]{\includegraphics{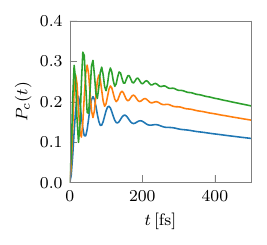}}
     \caption{Analysis of loss mechanisms in the polaritonic aggregate on changing relative loss rates and couplings.}
     \label{fig:tau_loss}
 \end{figure}

As a final bit of exploration, let us see how the transport efficiency of
exciton drain gets affected vis-\`a-vis the change in cavity properties ---
monomer-cavity coupling, $\Omega$, and cavity leakage rate, $\gamma_c\,(=1/T_c)$.
It is more meaningful to vary these parameters relative to their monomeric
counterparts --- $h$ and $\gamma_3\,(=1/T_3)$. Figure~\ref{fig:tau_loss}~(a) and
(b) show the change of $L_3^\infty$ and $\tau_3$ respectively as functions of
$\gamma_c/\gamma_3$ for different cavity couplings $\Omega/h$ keeping the
environment at a constant reorganization energy of $\lambda_0$. First consider
the amount of excitation extracted from the third monomer, $L_3^\infty$, shown
in Figure~\ref{fig:tau_loss}~(a). As a function of $\gamma_c/\gamma_3$ it goes
down monotonically because for the same $\Omega/h$, when $\gamma_c$
increases relative to $\gamma_3$, then more loss happens out of the cavity. What
is interesting is that for a particular $\gamma_c/\gamma_3$, the value of
$L_3^\infty$ decreases on increasing $\Omega/h$. To understand this better, we
plot the excitonic population of the third monomer and the cavity as a function
of time for the three values of $\Omega/h$ at a constant $\gamma_c/\gamma_3 =
0.25$ in Figure~\ref{fig:tau_loss}~(c) and (d) respectively. Notice that as
$\Omega/h$ increases, the amount of population buildup on the cavity increases,
but the accumulation on the third monomer decreases. This correspondingly
means that there is comparatively less population to be extracted out of
$\ket{3}$, leading to a decrease in $L_3^\infty$.

In Figure~\ref{fig:tau_loss}~(b), we see that the $\tau_3$ values for a constant
$\Omega/h$ decrease monotonically with $\gamma_c/\gamma_3$ as well. Probably
this is related to having less amount of exciton extracted from the molecular
drain ($L_3^\infty$). We conclude our discussion of this particular problem by
pointing out a surprising observation for future exploration: the $\tau_3$
curves for different $\Omega/h$ in Figure~\ref{fig:tau_loss}~(b) all intersect
at a point. This raises several very interesting questions: Why do we have this
point of intersection? Do polaritonic aggregates of different sizes all have
similar points of intersection? How does it change with changing bath on each
site? These interesting problems will be dealt with in a future publication
using the non-Hermitian state-to-state analysis technique developed in the
current letter.

In conclusion, we have described a new way for analyzing the end-point or trap
specific  efficiency of a multi-trap transport aggregate described by a
non-Hermitian Hamiltonian. This method is numerically exact when paired with
exact dynamics and makes no additional approximations. While here we have
explored the time-scales of transport, there can be several cases, where the
extraction process may not be appropriately described by a curve with a single
time scale. For such cases, the current non-Hermitian state-to-state analysis
technique can yield the full dynamics of extraction. In addition, being a
generalization of the state-to-state analysis
technique~\cite{boseImpactSolventStatetoState2023}, this method also allows us
to explore the exact pathways of transport under these leakages. In the examples
shown, we demonstrated how the non-Hermitian state-to-state method can be used
to understand transport in a polaritonic aggregate, where the exciton is
extracted from one of the molecules and the cavity is lossy. Of course, in such
a case, any loss of the excitation as a photon through the cavity does not count
towards transport. Using our non-Hermitian state-to-state method, we are able to
partition the total loss into the loss through the cavity and that through the
molecule. In the process, we reveal a wealth of extremely rich physics. While
the examples showed here used exact dynamics generated using path integrals, one
could as well use approximate semiclassical or perturbative methods to generate
the dynamics. This method promises to be an extremely powerful analysis tool for
understanding the dynamics of complex systems.

\bibliography{library}
\end{document}